\documentclass[showpacs,aps,prd,nofootinbib,floatfix,amsmath,amssymb]{revtex4}
\usepackage{graphicx}

\begin{document}

\makeatletter
\newbox\slashbox \setbox\slashbox=\hbox{$/$}
\newbox\Slashbox \setbox\Slashbox=\hbox{\large$/$}
\def\pFMslash#1{\setbox\@tempboxa=\hbox{$#1$}
  \@tempdima=0.5\wd\slashbox \advance\@tempdima 0.5\wd\@tempboxa
  \copy\slashbox \kern-\@tempdima \box\@tempboxa}
\def\pFMSlash#1{\setbox\@tempboxa=\hbox{$#1$}
  \@tempdima=0.5\wd\Slashbox \advance\@tempdima 0.5\wd\@tempboxa
  \copy\Slashbox \kern-\@tempdima \box\@tempboxa}
\def\FMslash{\protect\pFMslash}
\def\FMSlash{\protect\pFMSlash}
\def\miss#1{\ifmmode{/\mkern-11mu #1}\else{${/\mkern-11mu #1}$}\fi}
\makeatother

\newcommand{\zp}    {\mbox{$Z ^{\prime}$}}
\newcommand{\zpsm}    {\mbox{$Z ^{\prime}_{\rm SM}$}}

\title{Effects of an extra $Z'$ gauge boson on the top quark decay $t\to c\gamma$}
\author{A. Cordero--Cid}
\affiliation{Facultad de
Ciencias F\'\i sico Matem\'aticas, Benem\'erita Universidad
Aut\'onoma de Puebla, Apartado Postal 1152, Puebla, Pue., M\' exico}
\author{G. Tavares--Velasco}
\author{J. J. Toscano}
\affiliation{Facultad de
Ciencias F\'\i sico Matem\'aticas, Benem\'erita Universidad
Aut\'onoma de Puebla, Apartado Postal 1152, Puebla, Pue., M\'
exico}

\date{\today}

\begin{abstract}
The effects of an extra $Z'$ gauge boson with family nonuniversal
fermion couplings on the rare top quark decay $t\to c\gamma$ are
first examined in a model independent way and then in the minimal
331 model. It is found that the respective branching fraction is at
most of the order of $10^{-8}$ for $m_{Z'}=500$ GeV and dramatically
decreases for a heavier $Z'$ boson. This results is in sharp
contrast with a previous evaluation of this decay in the context of
topcolor assisted technicolor models, which found that $B(t\to c
\gamma)\simeq 10^{-6}$ for $m_{Z'}=1$ TeV.
\end{abstract}


\maketitle

Several well-motivated extensions of the standard model (SM) predict
an extra neutral gauge boson $Z'$, which has both theoretical and
experimental motivations. For instance, it has been conjectured that
a $Z'$ boson can be helpful to describe the $Z$-pole data and atomic
parity violation. Much work has been gone in studying the
phenomenology of an extra $Z'$ boson \cite{REVIEW}, and experimental
data have been used to obtain lower bounds on its mass and the
$Z'-Z$ mixing angle $\theta$. It is not possible to obtain
model-independent bounds, but current data from collider
\cite{BCOLLIDER} and precision \cite{PREC} experiments yield
$m_{Z'}\gtrsim 500$ GeV and $\sin \theta \le 10^{-3}$ for standard
GUT models; more strong constraints on $m_{Z'}$, of the order of 1
TeV, are obtained in models with nonuniversal flavor gauge
interactions such as topcolor assisted technicolor \cite{TAT},
noncommuting extended technicolor \cite{NCET}, or the ununified
standard model \cite{USM}. Quite recently, the CDF collaboration
used the Fermilab Tevatron data to search for new massive neutral
particles decaying into lepton pairs \cite{Abulencia:2005nf}. As far
as new gauge bosons are concerned, they considered a new SM-like
$Z'$ boson ($Z'_{SM}$), the $Z'$ bosons from the $E_{6}$ model
($Z_{\chi}$, $Z_{\psi}$, $Z_{\eta}$, $Z_{I}$) and the one from the
littlest Higgs model ($Z_{H}$) . The following lower bounds were
obtained \cite{Abulencia:2005nf} for the masses of $Z'_{SM}$,
$Z_{\chi}$, $Z_{\psi}$, $Z_{\eta}$ and $Z_{I}$: 825, 690, 675, 720
and 615~GeV, whereas $m_{Z_H}$ was found to be larger than 885, 860
,805 and 725~GeV for the following values of the mixing parameter
cot$\theta_{H}$: 1.0, 0.9, 0.7 and 0.5, respectively.

If the couplings of the $Z'$ to the fermions are assumed to be
family universal, there is no flavor changing neutral current (FCNC)
effects mediated by this particle even if there is fermion flavor
mixing via the GIM mechanism. Nevertheless, it is  possible that the
$Z'$ couples nonuniversally to the fermions, thereby giving rise to
FCNCs. For instance, some theories require that the $Z'$ couplings
to the third family fermions are different than the ones to the
fermions of the first two families. This is the case of the $Z'$
predicted by the 331 model \cite{331model}, which is based on the
$SU(3)_L\times U(1)_X$ symmetry  and has attracted considerably
attention recently \cite{331pheno}.  This model is particularly
appealing due to its unique mechanism of anomaly cancelation:
instead of the usual cancelation between each fermion family, it is
necessary that all the three fermion families are summed over, which
automatically requires the existence of a number of families that is
multiple of 3. It has been conjectured that this may provide a hint
to the solution of the family number problem. There are thus good
motivations for an extra $Z'$ boson, with a mass in the range 500
GeV--1 TeV, which couples nondiagonally to the fermions.

Although the phenomenology of a $Z'$ gauge boson is interesting by
itself, so is the study of FCNC effects as they would be a hint of
new physics because of their large suppression in the SM. This class
of effects has been considerably studied in the literature via some
rare decay modes of the top quark. The interest in the top quark
phenomenology stems from its large mass, which has led to some
belief that new physics effects are more likely to show through
processes involving this particle. Even more, the copious production
of top quark pairs at the CERN large hadron collider (LHC) will
allow us to examine several top quark properties and  some of its
rare decays. It is thus very interesting to consider the virtual
effects of a $Z'$ gauge boson on rare top quark decays. Among this
class of processes, the decay $t\to c \gamma$ has been analyzed both
in the SM and several of its extensions
\cite{Eilam:1990zc,Couture:1994rr}. While the SM predicts that the
$t\to c \gamma$ branching fraction is of the order of $10^{-10}$
\cite{Eilam:1990zc}, other models can enhance it up to $10^{-5}$
\cite{Couture:1994rr}.

In this brief report we will consider the contribution of the $Z'$
gauge boson to the $t\to c\gamma$ decay. It has been claimed
recently \cite{Yue:2003wd} that $B(t\to c \gamma)$ is at the
$10^{-6}$ level in topcolor assisted technicolor models for
$m_{Z'}=1$ TeV, which would allow the detection of this decay
channel at future particle colliders. Prompted by this assertion, we
will evaluate this decay in the context of some other models. Rather
than concentrating on a particular model, we will take the approach
of effective theories and consider an effective interaction $Z'qq'$.
We then present general expressions for the one-loop calculation.
This would allow us to have an estimate of the order of magnitude of
the $t\to c\gamma$ decay rate in a model-independent fashion, and
will be useful to assess whether the $Z'$ virtual effects have the
chance of becoming evident in future particle colliders through this
decay mode.

We will consider an effective interaction of the form

\begin{equation}
{\cal L}^{Z'q_iq_j}=\frac{ig'}{2c_W}\sum_{i,\,j}
\bar{q}_i\,\gamma_\mu\left(\xi_V^{ij}+\xi_A^{ij}\gamma^5\right)q_j
Z'^{\mu},\label{interaction}
\end{equation}
where the coefficients $\xi_V^{ij}$ and $\xi_A^{ij}$ contain the
information concerning any specific model. We refrain from
discussing the mechanism of generation of the FCNCs in models in
which the $Z'$ boson has family nonuniversal couplings, but a
discussion along this line can be found in Ref. \cite{Langacker}. We
first present results for the decay $t\to c\gamma$ in a
model-independent fashion and then examine it in the context of the
minimal 331 model. As a by-product, we reevaluate the $t\to c\gamma$
decay in the scenario posed by topcolor assisted technicolor models
and compare our result with previous ones.


\begin{figure}[!tbh]
\centering
\includegraphics[width=2.7in]{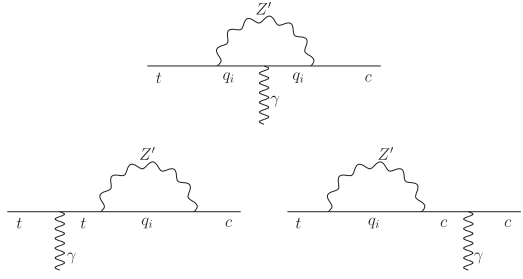}
\caption{\label{FeynDiag}Generic Feynman diagrams depicting the
contribution of a $Z'$ gauge boson to the top quark decay $t \to c
\gamma$.}
\end{figure}

The $Z'$ contribution to the top quark decay $t\to c \gamma$
proceeds through the Feynman diagrams shown in Fig. \ref{FeynDiag}.
Electromagnetic gauge invariance restricts the amplitude of this
decay to have the form

\begin{equation}
\label{amp} {\cal M}(t\to c\gamma)=\frac{i\,e\, g^2\epsilon^*_\alpha
k_\mu}{2^6\,c_W^2\,\pi^2\,m_t}\bar{u}_c(p_2)\sigma^{\alpha\mu}\left(F_1
+F_2\gamma_5\right)u_t(p_1),
\end{equation}
where $k_\alpha$ is the four-momentum of the photon. We will
calculate the $t\to c\gamma$ decay using the interaction
(\ref{interaction}) and verify that the resultant amplitude obeys
Eq. (\ref{amp}), i.e. that the coefficients of the gauge
noninvariant terms $\gamma^\alpha$ and $\gamma^\alpha\gamma^5$
vanish. We obtain in the massless $c$ quark limit:

\begin{equation}
F_1=\sum_{q=t,\,c}\left(\epsilon_A^{qc}
\epsilon_A^{tq}A_1(m_q)+\epsilon_V^{qc}
\epsilon_V^{tq}A_2(m_q)\right), \label{F1}
\end{equation}

\begin{equation}
F_2=-\sum_{q=t,\,c}\left(\epsilon_V^{qc}
\epsilon_A^{tq}A_1(m_q)+\epsilon_A^{qc}
\epsilon_V^{tq}A_2(m_q)\right),\label{F2}
\end{equation}
with $\epsilon^{ij}_{A,\,V}=g'/g\, \xi^{ij}_{A,\,V}$. We let the sum
run only over the $c$ and $t$ quarks as the associated Feynman
diagrams require only one FCNC vertex. We assume that those Feynman
diagrams including two FCNC vertices are more suppressed than those
containing just one of them. The $A_i$ factors are given by

\begin{eqnarray}
A_i(m_q)&=&\frac{1}{m_t^2m_{Z'}^2}\left(m_q^2\left(m_q^2-mt^2\right)+m_{Z'}^2\left(\left(m_q
\pm m_t\right)\left(m_q \pm 2\,m_t\right) -
2\,m_{Z'}^2\right)\right)\delta B_0(m_q)\nonumber\\
&-&\frac{m_q^2}{m_{Z'}^2}\left(m_q^2-m_t^2+2\,m_{Z'}^2\right)C_0(m_q)
-\frac{m_q\left(m_q\pm m_t\right)}{2\,m_{Z'}^2}-1,
\end{eqnarray}
where the $+$ ($-$) sign stands for $A_1$ ($A_2$), and

\begin{equation}
\delta B_0(m_q)=1+ \frac{\xi}{m_t^2}\,{\rm arccosh}\left(\frac{m_q^2
- m_t^2 + m_{Z'}^2}{2\,m_q\,m_{Z'}} \right) -
\frac{1}{2}\left(\frac{m_q^2 - m_{Z'}^2}{m_t^2} - \frac{ m_q^2 +
m_{Z'}^2}{m_q^2 - m_{Z'}^2} \right)
\log\left(\frac{m_q^2}{m_{Z'}^2}\right),
\end{equation}

\begin{equation}
C_0(m_q)=\frac{1}{m_t^2}\left[{\rm
Li}_2\left(1-\frac{m_{Z'}^2}{m_q^2}\right) -{\rm
Li}_2\left(\frac{2\,m_t^2}{\eta+\xi}\right)-{\rm
Li}_2\left(\frac{2\,m_t^2}{\eta-\xi}\right)\right],
\end{equation}
with $\xi^2=\left(m_q^2+m_t^2-m_{Z'}^2\right)^2- 4m_q^2m_t^2$ and
$\eta=m_q^2+m_t^2-m_{Z'}^2$. We have verified that ultraviolet
divergences cancel out. From (\ref{amp}), the decay width for $t\to
c\gamma$ follows easily:

\begin{equation}
\label{DecayWidth} \Gamma(t\to
c\gamma)=\frac{\alpha^3\,m_t\sum_i|F_i|^2}{2^{10}\,
\pi^2\,s_W^4\,c_W^4} \simeq 2.35 \times 10^{-7} {\rm GeV} \times
\sum_i|F_i|^2.
\end{equation}

\begin{figure}[!hbt]
\centering
\includegraphics[width=2.7in]{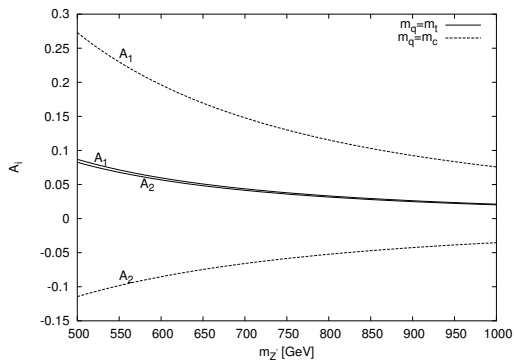}
\caption{\label{figa1a2}The  coefficients $A_i$  as functions of the
$Z'$ mass for $m_q=m_t$ and $m_q=m_c$.}
\end{figure}

Instead of evaluating the $t\to c\gamma$ branching ratio, it is
worth first evaluating the single contribution of an internal quark
to the $A_i$ coefficients. In this way we can get a rough estimate
of the order of magnitude of the branching ratio $B(t\to
c\gamma)\simeq\Gamma(t\to c\gamma)/\Gamma(t\to bW)$. In Fig.
\ref{figa1a2} we show the $A_i$ coefficients as functions of
$m_{Z'}$ for $m_q=m_t$ and $m_q= m_c$. We can see that both $A_1$
and $A_2$ are of the order of $10^{-1}$ for $m_{Z'}\simeq 500$ GeV
and decrease dramatically for a heavier $Z'$ boson. Even more, since
$A_2$ changes sign for $m_q=m_c$, there is the possibility of strong
cancelations when summing over the internal $c$ and $t$ quarks. We
can thus expect that the square $F_i$ form factors are of the order
of $10^{-2}$ times the square of some products of the
$\epsilon_{V,\,A}$ coefficients. In view of Eq. (\ref{DecayWidth})
we can conclude that even if the $\epsilon_{V,\,A}$ coefficients are
of the order of the unity and there is no strong cancelations, $B(t
\to c \gamma)$ could be of the order of $10^{-8}$ at most for
$m_{Z'}\simeq 500$ GeV.

Let us now analyze the scenario posed by the minimal 331 model
\cite{331model}. The details of this model are discussed for
instance in Ref. \cite{Perez}, here we will only present its most
relevant features to contextualize our results. In the
$SU(3)_L\times U(1)_X$ model the gauge interactions are non
flavor-universal since fermion generations are represented
differently  under the $SU_L(3)$ group. The leptons do not carry
quantum number $X$ and all the three generations are accommodated as
antitriplets of $SU_L(3)$. To cancel the $SU_L(3)$ anomaly, the same
number of fermion triplets and antitriplets is necessary, thereby
requiring two quark generations to be accommodated as triplets and
the other one as antitriplet. This is accomplished by introducing
three new exotic quarks, which are denoted by $D$, $S$, and $T$. The
electric charge of $D$ and $S$ is $-4/3 e$, whereas that of $T$ is
$5/3 e$. As far as the Higgs sector is concerned, it is comprised by
three $SU_L(3)$ triplets ($\phi_Y$, $\phi_1$ and $\phi_2$) and one
sextet ($\eta$). $\phi_Y$ is required to break $SU_L(3)\times
U_X(1)$ into $SU_L(2)\times U_Y(1)$, whereas the next stage of
spontaneous symmetry breaking (SSB) occurs at the Fermi scale and is
accomplished by the remaining triplets $\phi_1$ and $\phi_2$. The
scalar sextet $\eta$ is necessary to provide realistic masses for
the leptons \cite{FHPP}. In the gauge sector, the 331 model predicts
new particles: one pair of singly charged gauge bosons $Y^\pm$, one
pair of doubly charged gauge bosons $Y^{\pm\pm}$, and an extra
neutral boson $Z'$. These new gauge bosons and the exotic quarks get
their masses at the first stage of SSB. The charged gauge bosons
appear in an $SU_L(2)$ doublet with hypercharge $+3$ and carry two
units of lepton number, so they were dubbed bileptons. As for the
neutral fields, the symmetry breaking $SU_L(2)\times U_Y(1)\to
U_{Q}(1)$ yields the photon and two massive neutral gauge bosons $N$
and $N'$, which are linked to the mass eigenstates $Z$ and $Z'$ via
an unitary rotation. Since current constraints indicate that
$\sin\theta<10^{-3}$, it is reasonable to assume that the $Z-Z'$
mixing is negligible, in which case $Z$ ($Z'$) almost coincides with
$N$ ($N'$). In this scenario the flavor diagonal neutral currents
mediated by the photon and the $Z'$ boson can be written as
\begin{equation}
{\cal L}^{FD}=iQ_f \bar{f}\gamma_\mu f A^\mu+\frac{ig}{2c_W}
\bar{f}\gamma_\mu(g^f_{V\,Z'}-g^f_{A\,Z'}\gamma_5)fZ'^\mu.
\end{equation}
All the coefficients $g^f_{VZ'}$ and $g^f_{AZ'}$ are listed  in Ref.
\cite{Perez}, but we show the ones necessary for our calculation in
Table \ref{tabcoup}. As far as the FCNCs are concerned, they are
essentially mediated by the $Z'$ boson because the couplings of the
quarks to the $Z$ boson are proportional to $\sin\theta$. This class
of effects arises from the left-handed sector only and is a result
of the different $X$ quantum number assignments existing between the
fermion families. Assuming that there is no $Z-Z'$ mixing, the FCNC
Lagrangian for the up sector can be written as

\begin{equation}
\mathcal{L}^{FC}=\frac{ig}{2c_W}\, Z'^\mu\,
V^*_{3i}V_{3j}\overline{U}_i\gamma_\mu P_LU_j,
\end{equation}
with $\delta_L=\frac{2}{\sqrt{3}}\,\frac{c^2_W}{\sqrt{1-4s^2_W}}$,
$V_{ab}$ is the unitary matrix linking gauge states to mass
eigenstates, and $U_a=u,\,c,\,t$. In the right-handed sector there
is no FCNC as these fermions transform identically. The $V_{ab}$
matrix elements has be constrained via some FCNC processes
\cite{Perez}.

\begin{table}[!hbt]
\caption{\label{tabcoup} Diagonal couplings of the $Z'$ boson to up
quarks.}
\begin{tabular}{ccc}
\hline\hline $q$&$\epsilon_{V}$&$\epsilon_{A}$\\
\hline $t$&$\frac{1+4s_W^2}{2\sqrt{3}c_W^2\sqrt{1-4s_W^2}}$
&$\frac{\sqrt{1-4s_W^2}}{2\sqrt{3}c_W^2}$\\
$u,\,c$&$-\frac{1-6s^2_W}{2\sqrt{3}c^2_W\sqrt{1-4s^2_W}}$
&$-\frac{1+2s^2_W}{2\sqrt{3}c^2_W\sqrt{1-4s^2_W}}$\\
\hline\hline
\end{tabular}
\end{table}

\begin{figure}[!hbt]
\centering
\includegraphics[width=2.7in]{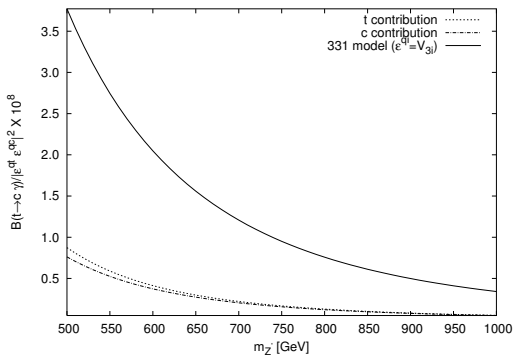}
\caption{\label{figbr331}Individual contributions of the $c$ and $t$
quarks to $B(t \to c \gamma)$ as a function of $m_{Z'}$ in $Z'$
models with left-handed couplings to up quarks. The solid line is
for the total contribution in the minimal 331 model.}
\end{figure}

Once the main features of the model were described, we proceed to
calculate the corresponding $t\to c\gamma$ branching ratio, which is
shown in Fig. \ref{figbr331} as a funcion of $m_{Z'}$. From that
Figure one can conclude that $B(t\to c\gamma)< 10^{-8}$ for
$m_{Z'}>500$ GeV. The inequality stems from the fact that the mixing
matrix elements $V_{3c}$ and $V_{3t}$ are smaller than the unity and
so is the coefficient $|V_{3c}V_{3t}|^2$.

The $Z'$ contribution to  $t\to c\gamma$ was previously calculated
in the context of topcolor-assisted technicolor (TC2) models
\cite{Yue:2003wd}. Those authors claim that, in that class of
models, $B(t\to c\gamma)$ ranges between $1.3\times10^{-6}$ and
$1.7\times10^{-7}$ for 1 TeV $\leq m_{Z'}\leq $ 2 TeV. Since our
previous findings seems to be in contradiction with that result, it
is convenient to reevaluate $B(t\to c\gamma)$ in the scenario
discussed in \cite{Yue:2003wd}. The flavor diagonal and flavor
changing couplings of the $Z'$ boson relevant for our calculation
are

\begin{equation}
\label{TCLFD} {\cal L}^{FD}_{Z'}=\frac{1}{6}i
g_1\cot\theta'\left(\bar{t}_L\gamma_\mu t_L+4\bar{t}_R\gamma_\mu
t_R\right)Z'^\mu,
\end{equation}

\begin{equation}
\label{TCLFC} {\cal L}^{FC}_{Z'}=-\frac{1}{6}i g_1
K_{tc}\left(\bar{t}_L\gamma_\mu c_L+4\bar{t}_R\gamma_\mu
c_R\right)Z'^\mu,
\end{equation}
where $g_1$ is the hypercharge coupling constant,
$\tan\theta'=g_1/\sqrt{4\pi k_1}$, and $K_{tc}$ is a flavor mixing
factor. It is required that $\tan\theta'\ll 1$ to have top quark
condensation rather than $b\bar{b}$ condensation \cite{Yue:2003wd}.

It is straightforward to cast Eqs. (\ref{TCLFD}) and (\ref{TCLFC})
in the form of Eq. (\ref{interaction}). After identifying the
coefficients that enter into Eqs. (\ref{F1}) and (\ref{F2}),
numerical evaluation yields $B(t\to c\gamma)\simeq 10^{-11}$ for
$m_{Z'}= 500$ GeV and the same set of parameters used in
\cite{Yue:2003wd}, i.e. $K_{tc}=0.8$ and $k_1=1$. For a heavier $Z'$
boson $B(t\to c\gamma)$ is negligibly small. This result is several
orders of magnitude below than the one found in Ref.
\cite{Yue:2003wd}. Unfortunately those authors did not report the
analytical result used for their calculation and we are unable to
compare it with ours.

Finally, we would like to consider the scenario in which the $Z'$
boson has only left-handed couplings to up quarks. This means that
we will take $\epsilon_{V}=\epsilon_{A}=\epsilon$ for both diagonal
and nondiagonal $Z'$ couplings. In Fig. \ref{figbr331} we have
plotted the single contributions of those loops carrying the same
internal quark to the $t \to c\gamma$ branching ratio. These results
show that $B(t\to c \gamma)$ is unlikely to be above the $10^{-8}$
level unless the $\epsilon$ coefficient is much larger than unity.

In closing, we would like to remark that the $Z'$ contribution to
the $t\to c\gamma$ decay is highly suppressed and it is unlikely
that this class of effects is at the reach of future particle
colliders. Our result is consistent with previous evaluations of
this decay in other models \cite{Eilam:1990zc}, which found, after
considering the current bounds, that the main contributions arise
from scalar particles rather than gauge bosons. As for the
contribution to the decays $t\to c g$ and $t\to c Z$, explicit
evaluation of the former shows that it can be of the order of
$10^{-6}$ at most for $m_{Z'}=500$ GeV and coupling constants of the
order $O(1)$ [we only need to make the replacement $\alpha\to
4/3\alpha_s$ in Eq. (\ref{DecayWidth}) to obtain $\Gamma(t\to cg)$],
while the latter is expected to be more suppressed due to
kinematics.

\acknowledgments{We acknowledge support from PROMEP and Conacyt
(M\'exico) under grant U44515-F.

\end{document}